\documentclass[11pt]{article}

\usepackage[margin=1in]{geometry}
\usepackage{graphicx}
\graphicspath{{Calibrating_trust/}}   
\usepackage{amsmath}
\usepackage{booktabs}
\usepackage{hyperref}
\usepackage{caption}
\usepackage{subcaption}
\usepackage{float}       
\usepackage{authblk}
\usepackage{placeins}    
\usepackage{threeparttable}   
\usepackage{tabularx}         
\usepackage{threeparttable}   
\usepackage{pdflscape}        
\usepackage{adjustbox}        

\title{\textbf{What Shapes User Trust in ChatGPT?\\
A Mixed-Methods Study of User Attributes, Trust Dimensions, Task Context, and Societal Perceptions among University Students}}

\author[1,2]{Kadija Bouyzourn}
\author[1]{Alexandra Birch}
\affil[1]{University of Edinburgh}
\affil[2]{KU Leuven}
\date{}    

\begin{document}
\maketitle

\begin{abstract}
This mixed-methods inquiry examined four domains that shape university students’ trust in ChatGPT: user attributes, seven delineated trust dimensions, task context, and perceived societal impact. Data were collected through a survey of 115 UK undergraduate and postgraduate students and four complementary semi-structured interviews. Behavioural engagement outweighed demographics: frequent use increased trust, whereas self-reported understanding of large-language-model mechanics reduced it. Among the dimensions, perceived expertise and ethical risk were the strongest predictors of overall trust; ease of use and transparency had secondary effects, while human-likeness and reputation were non-significant. Trust was highly task-contingent; highest for coding and summarising, lowest for entertainment and citation generation, yet confidence in ChatGPT’s referencing ability, despite known inaccuracies, was the single strongest correlate of global trust, indicating automation bias. Computer-science students surpassed peers only in trusting the system for proofreading and writing, suggesting technical expertise refines rather than inflates reliance. Finally, students who viewed AI’s societal impact positively reported the greatest trust, whereas mixed or negative outlooks dampened confidence. These findings show that trust in ChatGPT hinges on task verifiability, perceived competence, ethical alignment and direct experience, and they underscore the need for transparency, accuracy cues and user education when deploying LLMs in academic settings.
\end{abstract}

\section{Introduction}
Large Language Models (LLMs) such as ChatGPT are reshaping human-AI interaction by accelerating content creation, knowledge retrieval and decision support. Yet their tendency to fabricate content, hallucinate citations or deliver biased answers threatens reliability \cite{shen2023reliability}. This capability-accuracy tension makes trust central to responsible use; effective designs must reflect user and stakeholder needs \cite{yoo2024}. Trust is not simply believing AI outputs; it is deciding when, how and why to rely on them across tasks \cite{hoff2015}.

Although trust is widely viewed as essential, existing research is conceptually scattered and overly general \cite{benk2025}. Reviews of conversational AI show wide variation in how trust is defined and measured \cite{ng2025}, and Human-Centred AI studies reveal similar inconsistencies \cite{capel2023}.

This fragmentation calls for context-specific, empirically grounded models \cite{benk2025}. Comparable multidimensional adoption frameworks exist for other emerging technologies \cite{acheampong2019}. Consistent with behavioural theory \cite{ajzen1991}, trust reflects attitudes, perceived norms and perceived control that shape reliance. Addressing gaps in educational-AI research \cite{fu2024}, we study ChatGPT among university students, a group where trust closely tracks AI attitudes \cite{ramirez2024}. We analyse how user traits, task types and perceived performance shape trust, since understanding these links is vital for system design and usability \cite{eiband2018}.

Frameworks such as Hoff and Bashir’s \cite{hoff2015} dispositional-situational-learned model show that trust evolves with context and experience. Automation-bias studies reveal how perceived reliability drives over- or under-reliance \cite{dzindolet2003}. For LLMs, transparency is key to trust calibration \cite{zerilli2022patterns}, yet few empirical studies map how trust varies by task, user profile and system traits.

We therefore examine trust in ChatGPT across four dimensions: (1) user characteristics, (2) specific trust attributes (e.g., expertise, predictability), (3) task type (translation, summarisation, coding) and (4) perceived societal and ethical impact. This multidimensional lens clarifies when and why users trust generative AI and informs responsible integration into knowledge work. Using a mixed-methods design, we pair a survey ($n = 115$) covering demographics, usage and trust with interviews that unpack participants’ reasoning, yielding a nuanced view of how users build and calibrate trust.

Results show that behaviour drives trust. Frequent users trust more, whereas those with deeper technical understanding are more cautious, mirroring findings that usage intent, perceived safety and AI competence shape attitudes \cite{katsantonis2024, delcker2024}. Trust also varies by task: participants trust ChatGPT for summarisation, coding and information retrieval, but not for citing references. Nonetheless, perceived referencing ability strongly predicts overall trust, suggesting over-reliance on confident output. Among trust attributes, expertise, predictability, ease of use and transparency matter most; human-likeness and general reputation matter least. Ethical concerns such as bias, privacy and academic misuse are pervasive, and results suggest that trust is influenced by broader societal awareness. 
\section*{Background}

AI’s rapid spread makes user trust pivotal for acceptance, adoption and governance \cite{bach2024}. Trust goes beyond technical accuracy; it is a multidimensional relationship shaped by transparency, adaptability and perceived agency \cite{hoff2015, schwartz2022}. Ethically, the AI4People framework calls for transparency, accountability and respect for human agency, urging designs that empower users while curbing over-reliance and opacity \cite{floridi2018}. Systems should enable critical review and bias mitigation, promoting calibrated trust \cite{zerilli2022patterns}. Empirical work confirms that transparency directly boosts trust and acceptance \cite{wanner2022}.

Yet research on AI trust remains fragmented and lacks shared concepts or context-specific models \cite{benk2025, lukyanenko2022}. In chatbots, reviews expose inconsistent definitions and measures \cite{ng2025}, and main-path analysis shows dispersion across disciplines \cite{henrique2024}. Against this backdrop, we examine ChatGPT, a widely used LLM whose content generation, language support and educational promise eclipse rule-based tools \cite{shen2023radiology}. Nonetheless, ethical, transparency and reliability concerns remain acute, especially in high-stakes contexts \cite{shen2023reliability}. Trustworthy AI must clarify its limits and ground the user relationship in competence and responsibility \cite{floridi2018}. This interdisciplinary lens frames ChatGPT’s specific opportunities and risks.

\section*{Conceptualising Trust}

Trust in AI is a multidimensional construct spanning psychology, sociology and human-computer interaction. Classic definitions cast it as confidence under uncertainty \cite{mayer1995}, a mechanism for reducing social complexity \cite{luhmann2017} and a cognitive aid when facing limited information \cite{gambetta1988}. For AI, trust is best understood as epistemic trust: reliance on the system strictly as an information source \cite{alvarado2023}.

Hoff and Bashir’s dispositional-situational-learned model highlights trust’s fluid nature, which evolves through interaction \cite{hoff2015, lee2004}, and balanced calibration averts both misuse and underuse \cite{axelrod1981}. As perceptions must align with actual capability \cite{hoff2015, nist2023rmm}, trust grows slowly but erodes quickly, reflecting performance outcomes and competence signals \cite{yang2016adjust, yang2023dynamics} as users continually update reliability estimates \cite{ajzen1975}.

Familiarity sharpens this calibration: technically literate or experienced users gauge system limits more accurately \cite{bach2024, foehr2020}, whereas novices often over- or under-trust, underscoring the value of user-centred design \cite{elkins2013, klumpp2019}. Users also triangulate outputs against prior knowledge and external sources \cite{rowley2013}, a critical safeguard for large language models whose responses are not always verifiable.

Accordingly, we define trust in ChatGPT as calibrated reliance shaped by user attributes, task context and perceived system qualities. Adopting Hoff and Bashir’s dynamic view, we examine both cognitive and affective dimensions and assess trust at the task level, focusing on how features such as transparency and predictability influence reliance. This approach aligns with recent calls for context-specific, multidimensional frameworks \cite{benk2025, choudhury2023}.

\section*{Dimensions of Trust in AI-Driven Systems}

Trust in AI rests on several intertwined dimensions that shape user evaluation and behaviour. Transparency is primary: by exposing an AI system’s logic, limitations and reasoning, it encourages critical engagement rather than blind reliance \cite{zerilli2022patterns}. Well-timed disclosure of weaknesses can raise calibrated trust and even enhance performance \cite{rieger2025}, yet explanations do not always boost confidence; poorly framed details may erode it \cite{schmidt2020}. Performance cues, error rates, confidence scores and similar metrics, support calibration by steering users between over-trust and undue scepticism \cite{dietvorst2020, dzindolet2003}. Real-time feedback further refines situational trust, helping users form accurate mental models of system behaviour \cite{nist2023rmm, ehsan2020}. Together, these mechanisms align with the NIST AI 600-1 \cite{nist2024genai} guidance on balanced engagement with generative models such as ChatGPT.

Interface design also matters. Humanlike cues, such as responsiveness, conversational tone and perceived expertise, can increase comfort, but functional qualities such as reliability and clarity are stronger predictors of trust \cite{nordheim2019}. Specific anthropomorphic elements, for example a consistent ‘personality,’ may ease interaction \cite{foehr2020}; however, they must be paired with clear statements of limitation to avoid automation bias \cite{hoff2015} and the misplaced confidence it breeds \cite{shen2023reliability}. Accordingly, NIST AI 600-1 \cite{nist2024genai} recommends coupling humanlike design with transparency.

Perceived competence, belief that a system delivers accurate, consistent results, becomes critical in high-stakes domains such as healthcare or law, where minor errors can dissolve trust \cite{shen2023radiology, shen2023reliability, manzey2012}. Here, predictability and reliability reinforce confidence \cite{muir1987, madhavan2007}.

Guided by the seven-dimension framework of Choudhury and Shamszare, namely, expertise, predictability, transparency, human-likeness, ease of use, risk and reputation, our study analyses how each dimension shapes user trust in ChatGPT. Viewing trust as dynamic and context-bound, this lens enables a nuanced assessment rather than a single global metric.

\section*{Task-Specific Trust and Contextual Considerations}

Trust in AI varies with task complexity and stakes. Users show higher confidence in binary-correct tasks such as coding than in domains where errors have grave consequences, notably healthcare or law \cite{almarazlopez2023, zerilli2022patterns, zerilli2019double}. Task-specific feedback further calibrates reliance, especially in high-stakes settings, by guiding users to validate outputs through triangulation, or consulting multiple sources \cite{nist2024genai, rowley2013}. Clear scoping also matters: AI that performs narrowly defined functions, process automation, cognitive insights or cognitive engagement, earns more trust when it delivers consistent, predictable results \cite{davenport2018}.

\section*{Socio-Ethical Considerations}

Privacy, accountability and fairness are pillars of trustworthy AI \cite{bach2024}. Perceived bias or unethical conduct quickly undermines confidence, making transparent, value-aligned design imperative \cite{binns2017}. Over-trust, uncritical acceptance of AI output, poses acute risks in high-stakes domains and is well documented as automation bias \cite{shen2023reliability, shen2023radiology, parasuraman2010}. Accordingly, Choudhury and Shamszare \cite{choudhury2023} call for strong safeguards against misuse and discrimination.

Policy guidance reinforces these demands. The EU’s AI HLEG and consecutive NIST frameworks position transparency and accountability as foundation stones of trustworthy AI, insisting on mechanisms that surface limitations and rectify errors \cite{ec2019, nist2023rmm, nist2024genai}. Grant et al. \cite{grant2025} add that ethical transparency must include procedural fairness to protect decision-subjects’ rights. Echoing this stance, the AI4People principles advocate transparency, accountability and user empowerment \cite{floridi2018}. Russo et al. \cite{russo2024} integrate epistemic and ethical dimensions, urging evaluation of the entire AI lifecycle, while Shneiderman \cite{shneiderman2022} synthesises these threads in a human-centred agenda that embeds safety, agency and values directly into system design.
\section{Methods}

This study examines how users develop trust in ChatGPT through four research questions. RQ1 investigates how user characteristics (such as field of study, familiarity and usage frequency) influence trust. RQ2 explores how trust dimensions (e.g.\ expertise, predictability, transparency) interact. RQ3 looks at how trust varies across different task types. RQ4 considers how perceptions of AI’s societal impact relate to trust. A mixed-methods approach combines surveys to identify patterns and interviews to explore user reasoning, capturing both broad trends and individual insights.

\subsection{Participants}

The study involved 115 undergraduate and postgraduate students (aged 18–35) from the University of Edinburgh, with participants drawn from various academic disciplines, including a strong representation from computer science. Recruitment was conducted through university channels such as course announcements and social media, with voluntary participation and informed consent ensuring ethical compliance. Demographic data collected included age, degree level, field of study, AI familiarity and frequency of ChatGPT use. Most participants reported moderate to high familiarity with ChatGPT and other large language models, reflecting their growing presence in higher education. The disciplinary diversity provided a strong basis for comparing trust perceptions across fields.

\subsection{Data Collection Instruments}

\subsubsection{Quantitative Questionnaire}

The structured questionnaire (Appendix~B) included several sections to assess different aspects of trust in ChatGPT:
\begin{itemize}
  \item Demographics: Participants reported their age, degree level, field of study and behaviour related to AI technologies.
  \item Trust Dimensions: Using a 5-point Likert scale, participants rated statements related to expertise, predictability, transparency, human-likeness, ease of use, risk, and reputation. These dimensions were based on the framework by Choudhury \& Shamszare (2023), who identified them as key factors influencing trust in AI systems. Construct details and reliability scores are provided in Appendix~A.
  \item Task-Specific Trust: Participants rated their trust in ChatGPT across tasks such as summarising, coding, proofreading and idea generation, using a 5-point scale from no trust to complete trust.
  \item Societal Impact:\textbf{ }The final section addressed participants’ views on the ethical and societal implications of AI.
\end{itemize}

\subsubsection{Qualitative Interviews}

Four participants were interviewed, selected from those who expressed interest during the survey. The sub-sample included two computer science students (one undergraduate, one postgraduate) and two students from other disciplines (one undergraduate, one postgraduate). Interviews explored overall impressions of ChatGPT, task-specific trust, ethical concerns and the trust dimensions participants found most important.

A semi-structured format allowed for flexible, in-depth discussion. Interviews were audio-recorded, transcribed and thematically analysed to identify key themes and individual perspectives (Braun \& Clarke, 2006). All names were replaced with pseudonyms and identifying information was removed to protect anonymity.

\subsection{Ethical Considerations}

Ethical approval was granted by the University of Edinburgh Ethics Committee (reference 699585). All participants provided informed consent, with assurances of confidentiality and the right to withdraw at any time. To minimise potential biases, interview discussions were guided by a neutral script.


\section{Results}\label{sec:results}

This study included 115 participants, predominantly young adults, with a significant portion studying computer science. Participants reported high familiarity with ChatGPT, and frequent usage was common across the sample. Trust ratings for ChatGPT and LLMs ranged from moderate to high. To provide a comprehensive understanding of trust, quantitative results are complemented by qualitative insights gathered from interviews with a sub-sample of four participants. These qualitative findings are integrated thematically throughout the results to further elaborate on and contextualise the quantitative patterns.

\subsection{RQ1: How Do User Profiles and Experiences Influence Trust in ChatGPT?}

\begin{figure}[H]
  \centering
  \includegraphics[width=1\linewidth]{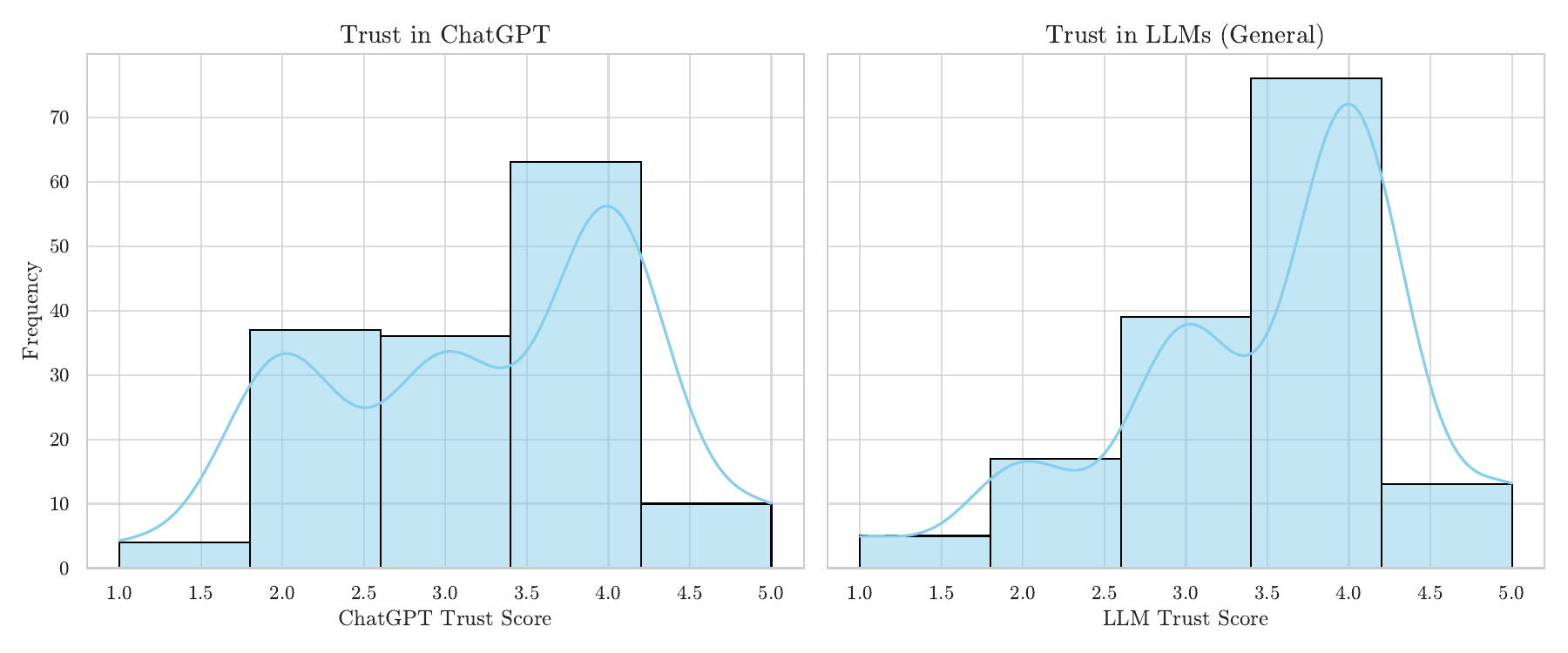}
  \caption{Distribution of trust scores for ChatGPT (left) and LLMs (right).
  \newline\textit{Note.} Scores were recorded on a 5-point Likert scale (1 = very low, 5 = very high).}
  \label{fig:trust_dist}
\end{figure}

This section examines how user characteristics influence trust in ChatGPT and other LLMs. Specifically, we assess whether demographic factors (such as field of study and degree level) and behavioural variables (such as familiarity, self-reported understanding, and frequency of use) affect trust levels. Participants rated their overall trust in ChatGPT and LLMs. As shown in Figure~\ref{fig:trust_dist}, trust ratings were generally high, with most reporting moderate to strong trust. Correlation analysis revealed a significant positive relationship between trust in ChatGPT and trust in LLMs ($\rho = 0.587$, $p < 0.001$), suggesting that users who trust LLMs overall also tend to trust ChatGPT.

Participants rated their trust in ChatGPT using a 5-point Likert scale. Descriptive statistics across demographic and behavioural variables are presented in Table~\ref{tab:demo_behav}. To examine variations in trust, participants were grouped according to key factors identified in the questionnaire. These groupings are detailed in Table~\ref{tab:demo_behav}.

\begin{table}[H]
\centering
\begin{threeparttable}
\caption{Descriptive statistics for ChatGPT trust by demographic and behavioural variables}
\label{tab:demo_behav}

\begin{tabularx}{\textwidth}{@{}p{2.9cm}p{3.4cm}Xcc@{}}
\toprule
\textbf{Category} & \textbf{Variable} & \textbf{Grouping} & \textbf{$N$} & \textbf{Mean (SD)}\\
\midrule
\multicolumn{5}{@{}l}{\textbf{Demographic}}\\[2pt]
 & Field of study          & Computer Science      & 77 & 3.29 (0.89)\\
 &                         & Other fields\textsuperscript{a} & 38 & 3.11 (0.95)\\[2pt]
 & Degree level            & Undergraduate         & 45 & 3.18 (0.89)\\
 &                         & Postgrad (taught)     & 36 & 3.14 (0.99)\\
 &                         & Postgrad (research)   & 34 & 3.38 (0.85)\\[6pt]

\multicolumn{5}{@{}l}{\textbf{Behavioural}}\\[2pt]
 & LLM familiarity         & Low–moderate\textsuperscript{b} & 57 & 3.25 (0.89)\\
 &                         & High                  & 58 & 3.21 (0.93)\\[2pt]
 & LLM understanding       & Not / uncertain       & 69 & 3.35 (0.94)\\
 &                         & Understands           & 46 & 3.06 (0.84)\\[2pt]
 & Usage frequency         & Infrequent\textsuperscript{c}   & 40 & 2.85 (0.98)\\
 &                         & Frequent              & 75 & 3.43 (0.81)\\
\bottomrule
\end{tabularx}

\begin{tablenotes}[para,flushleft]\footnotesize
\item \textit{Note.} $N = 115$. Trust was rated on a 1–5 Likert scale.%
\;%
\textsuperscript{a} Humanities, Social Sciences, STEM (non-CS).%
\;%
\textsuperscript{b} “Not familiar at all” to “Moderately familiar.”%
\;%
\textsuperscript{c} “About once per week” or less; “Frequent” = more than once per week.
\end{tablenotes}
\end{threeparttable}
\end{table}
To assess whether trust in ChatGPT varied by educational background, we compared trust scores across fields of study and degree levels. Research postgraduate students reported slightly higher trust, and Computer Science students had marginally higher scores than those from other fields (see Table~\ref{tab:demo_behav}). However, these differences were not statistically significant (Mann–Whitney $U = 1623.5$, $p = .31$; Kruskal–Wallis $H(2) = 1.65$, $p = .44$). These findings support RQ1 by showing that behavioural factors, especially usage frequency and perceived understanding, play a stronger role in shaping trust than familiarity alone. Regular interaction and a basic level of perceived knowledge may foster confidence, even when users lack deeper technical insight.

In contrast to demographic factors, behavioural variables revealed notable differences (see Table~\ref{tab:behav_sig}). Participants with limited or uncertain understanding of LLMs reported significantly higher trust in ChatGPT than those with more confident understanding, suggesting that greater technical awareness may foster more critical engagement. Familiarity with LLMs did not significantly affect trust. However, usage frequency emerged as a key factor: frequent users reported significantly higher trust than infrequent users. These results support RQ1, indicating that behavioural factors, especially usage frequency and perceived understanding, are more influential than familiarity in shaping trust.

\begin{table}[H]
\centering
\begin{threeparttable}
  \caption{Trust in ChatGPT by understanding, familiarity, and usage frequency}
  \label{tab:behav_sig}
  \begin{tabular}{lll}
    \toprule
    Variable & Group comparison & $p$-value\\
    \midrule
    LLM understanding & Understands vs.\ Not/uncertain & .049$^{*}$\\
    LLM familiarity   & High vs.\ Low                   & .868\\
    Usage frequency   & Frequent vs.\ Infrequent        & .001$^{*}$\\
    \bottomrule
  \end{tabular}
  \begin{tablenotes}[para,flushleft]\footnotesize
    * $p<.05$ (Mann–Whitney $U$ tests).
  \end{tablenotes}
\end{threeparttable}
\end{table}

To examine how behavioural variables predict trust in ChatGPT, an ordinal logistic regression was conducted with usage frequency, LLM familiarity, and LLM understanding as predictors (see Table~\ref{tab:logreg}). The results indicated that usage frequency and understanding significantly predicted trust. Specifically, more frequent use of ChatGPT was associated with higher trust, while greater understanding of how LLMs work was negatively associated with trust. Familiarity, however, was not a significant predictor. These findings contribute to RQ1 by demonstrating that behavioural engagement, particularly frequency of use, is a key factor in shaping trust in ChatGPT. In contrast, greater technical understanding may prompt more critical or cautious attitudes toward the system.

\begin{table}[H]
\centering
\begin{threeparttable}
  \caption{Ordinal logistic regression predicting trust in ChatGPT}
  \label{tab:logreg}
  \begin{tabular}{lrrrr}
    \toprule
    Predictor & B & SE & $z$ & $p$\\
    \midrule
    Usage frequency   &  1.14 & 0.38 &  2.97 & .003$^{*}$\\
    LLM familiarity   & –0.56 & 0.44 & –1.27 & .205\\
    LLM understanding & –0.90 & 0.45 & –2.01 & .044$^{*}$\\
    \bottomrule
  \end{tabular}
  \begin{tablenotes}[para,flushleft]\footnotesize
    * $p<.05$.
  \end{tablenotes}
\end{threeparttable}
\end{table}

\noindent\textbf{Interview Perspectives.}\;To complement the survey findings, we analysed interview data to understand how trust in ChatGPT develops over time and is shaped by users’ backgrounds, use cases, and familiarity. The interviews supported and elaborated on the quantitative trends, offering more nuanced perspectives on how users navigate trust in practice.

Ishaan, a postgraduate in biomedical AI, attributed his cautious trust to his technical understanding of machine learning: “\textit{I understand that it’s a model that’s based on […] text sources […] It might give you the right logic but not the correct execution.}” While he trusted ChatGPT for basic tasks like grammar correction, his confidence dropped for more complex tasks involving references:

\begin{quote}
I started building trust in ChatGPT for information gathering […] but then over time, when I started looking up the references and they were not existing, I realised I can’t trust it for more complex situations.\\
\hfill --- Ishaan, postgraduate CS student
\end{quote}

Tom, an undergraduate in computer science, also described a careful but evolving trust shaped by familiarity: “\textit{I have more of an understanding of how it functions […] I know not to trust it fully.}” His trust increased as he verified outputs and found them “\textit{accurate or close enough,}” but he remained cautious in high-stakes contexts: “\textit{I probably trust it more now than I did at the start […] but I still wouldn’t trust it fully for important tasks.}”

Among humanities students, Fernanda expressed scepticism early on, noting concerns about information credibility: “\textit{I know ChatGPT gathers information […] but I don’t know if it’s from credible sources.}” 

Ella, an undergraduate in modern languages, initially appreciated ChatGPT’s fluency but grew critical over time, citing repetitive structures and shallow content: “\textit{It sounds great if you skim it, but it kind of doesn’t really say anything […] It just uses the same sentence structures over and over.}” Her experience highlights how superficial or formulaic responses can erode trust, particularly in tasks involving critical or creative thinking.

\subsection{RQ2: How Do Dimensions of Trust Shape Overall Trust in ChatGPT?}

This section explores how dimensions of trust, namely expertise, predictability, transparency, human-likeness, ease of use, risk (framed as ethical compliance) and reputation, influence overall trust in ChatGPT, and how these perceptions vary across user groups. Participants rated their agreement with statements related to each dimension on a five-point Likert scale. Composite scores were calculated by averaging responses within each dimension, with reverse coding applied where necessary to ensure consistency (see Appendix A for construct details and reliability statistics). Overall, participants rated ChatGPT moderate to high on most trust dimensions. Scores were highest for \textit{expertise} (M = 3.8), \textit{ease of use} (M = 4.0), \textit{predictability} (M = 3.6), \textit{transparency} (M = 3.6), and \textit{reputation} (M = 3.5), reflecting generally positive perceptions across these dimensions. In contrast, \textit{human-likeness} (M = 3.1) and \textit{risk} (M = 3.2) showed greater variability, with flatter or more widely spread distributions. These results suggest that while functional and ethical aspects of ChatGPT are broadly trusted, participants hold more mixed or uncertain views about how human-like it seems or how well it handles risk and compliance.

\begin{figure}[H]
  \centering
  \includegraphics[width=.75\linewidth, trim=0 50 0 40, clip]{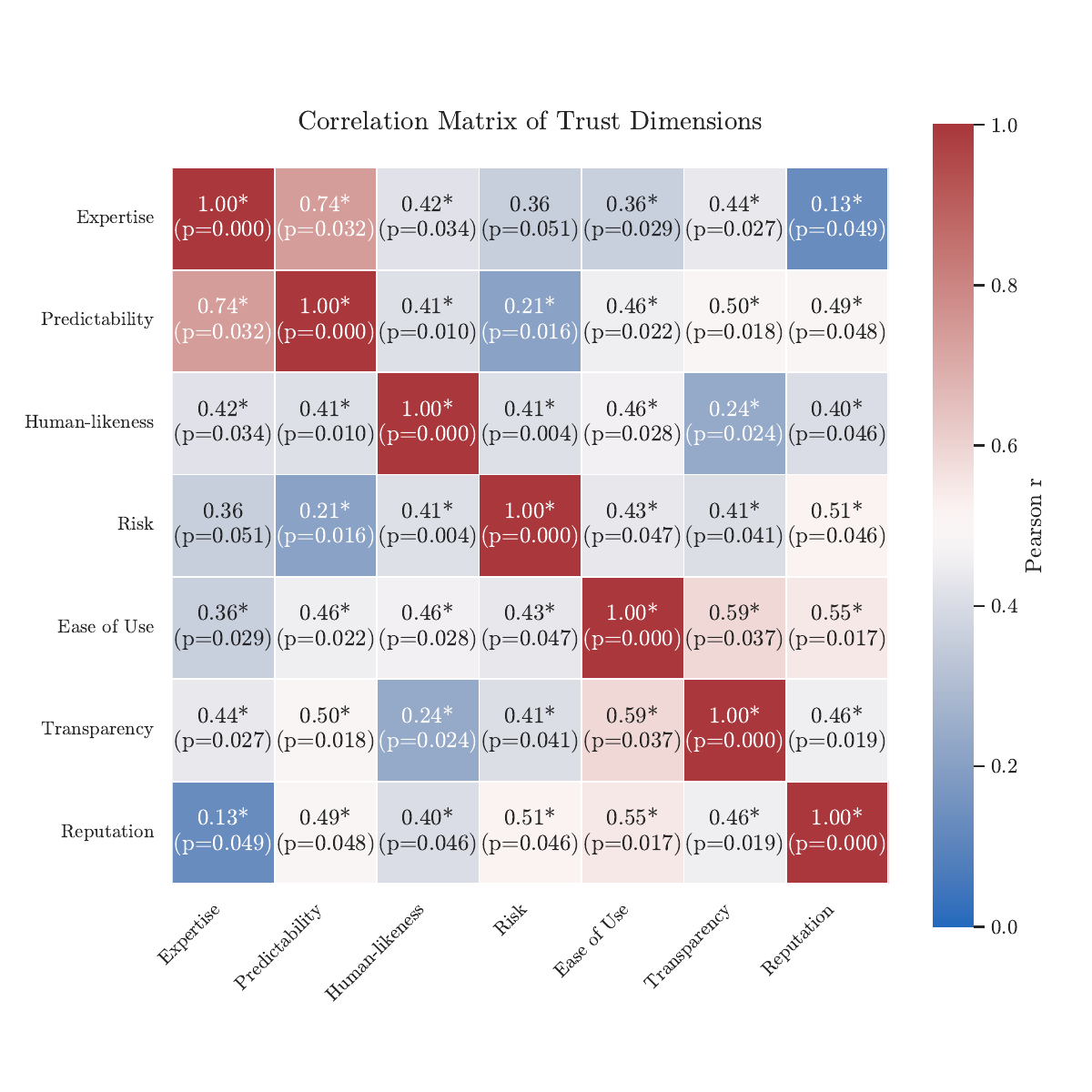}
  \caption{Spearman correlation matrix among trust dimensions.}
  \label{fig:corr_matrix}
\end{figure}

To examine how trust dimensions relate to one another, we analysed correlations between them (see Figure~\ref{fig:corr_matrix}). Several significant positive relationships emerged. For example, users who found ChatGPT easy to use tended to perceive it as less risky. Transparency was positively correlated with both expertise and ease of use, suggesting that clear responses enhance perceptions of competence and usability. Expertise and predictability were also moderately correlated, indicating that users who saw ChatGPT as knowledgeable also viewed it as reliable. These findings partially address RQ2 by suggesting that trust is shaped by a combination of perceived attributes such as competence, clarity, usability and reputation.

\FloatBarrier

To assess how each trust dimension relates to overall trust in ChatGPT, we conducted a correlation analysis using trust as the outcome variable (see Table~\ref{tab:dim_corr}). \textit{Expertise} showed the strongest correlation, followed by \textit{predictability}, \textit{ease of use}, and \textit{transparency}. This suggests that users trust ChatGPT more when it is perceived as competent, consistent, clear, and user-friendly. \textit{Reputation} also correlated with trust, though less strongly, indicating that personal experience may matter more than external perception. \textit{Human-likeness} was not significantly correlated, suggesting that mimicking human traits is not essential for trust. These findings partially address RQ2, showing that trust in ChatGPT is shaped more by functional and ethical attributes than by resemblance to human interaction.

\begin{table}[H]
\centering
\begin{threeparttable}
  \caption{Correlations between each trust dimension and overall trust in ChatGPT}
  \label{tab:dim_corr}
  \begin{tabular}{lrr}
    \toprule
    Trust dimension & $\rho$ & $p$\\
    \midrule
    Expertise        & 0.551 & $<.001^{*}$\\
    Predictability   & 0.286 & .002$^{*}$\\
    Human-likeness   & 0.135 & .150\\
    Risk             & 0.436 & $<.001^{*}$\\
    Ease of use      & 0.437 & $<.001^{*}$\\
    Transparency     & 0.443 & $<.001^{*}$\\
    Reputation       & 0.230 & .013$^{*}$\\
    \bottomrule
  \end{tabular}
  \begin{tablenotes}[para,flushleft]\footnotesize
    * $p<.05$ (two-tailed Spearman).
  \end{tablenotes}
\end{threeparttable}
\end{table}

To examine which dimensions best predict overall trust in ChatGPT, we ran a multiple linear regression using all seven trust dimensions as predictors (Table~\ref{tab:dim_reg}). The model explained 43.9\% of the variance in trust. Perceived expertise and risk were the strongest predictors, followed by ease of use and transparency. These results indicate that trust is primarily shaped by perceptions of competence, ethical alignment, clarity and usability. Human-likeness and reputation did not significantly predict trust, suggesting users value functional performance over anthropomorphic traits or external opinions.

\begin{table}[H]
\centering
\begin{threeparttable}
  \caption{Multiple regression of overall trust on the seven dimensions}
  \label{tab:dim_reg}
  \begin{tabular}{lrr}
    \toprule
    Predictor & B & $p$\\
    \midrule
    Expertise      & 0.34  & $<.001^{*}$\\
    Predictability & 0.03  & .755\\
    Human-likeness & –0.01 & .933\\
    Risk           & 0.19  & .017$^{*}$\\
    Ease of use    & 0.12  & .311\\
    Transparency   & 0.15  & .123\\
    Reputation     & –0.01 & .954\\
    \bottomrule
  \end{tabular}
  \begin{tablenotes}[para,flushleft]\footnotesize
    $R^{2} = 0.439$; * $p<.05$.
  \end{tablenotes}
\end{threeparttable}
\end{table}

\begin{figure}[H]
  \centering
  \includegraphics[width=.85\linewidth, trim=0 0 0 27, clip]{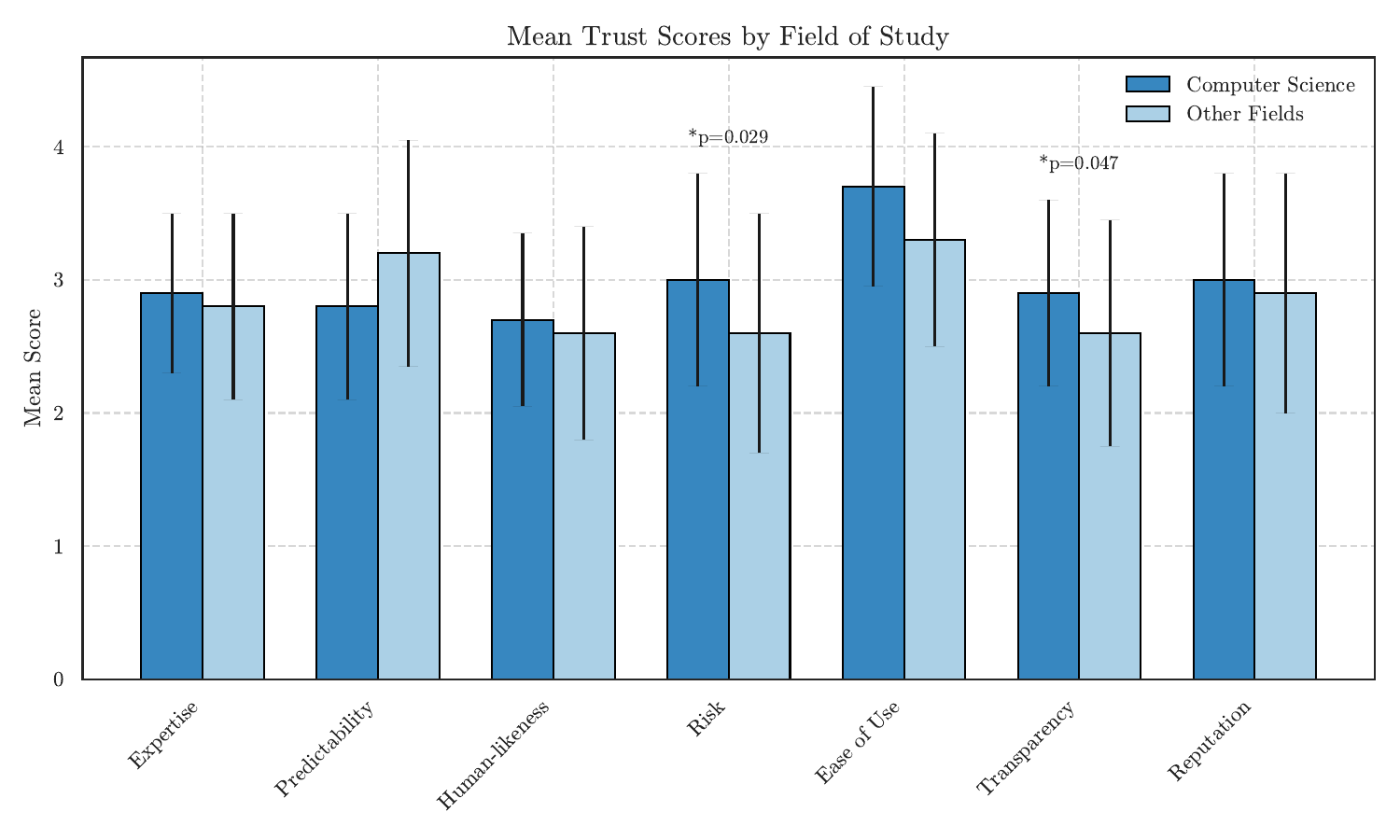}
  \caption{Mean trust-dimension scores by field of study.
  \newline\textit{Note.} Error bars = $\pm1$\,SE.}
  \label{fig:trust_by_field}
\end{figure}

Next, we compared trust-dimension scores across fields of study. As shown in Figure~\ref{fig:trust_by_field}, Computer Science students reported higher ratings on risk, ease of use, transparency and reputation, while students from other disciplines rated ChatGPT higher on predictability. 

\FloatBarrier

Statistically significant differences were found for transparency and risk, with Computer Science students scoring higher on both (Table~\ref{tab:dim_field}). These results suggest that participants with a technical background may perceive ChatGPT as more transparent and ethically compliant, whereas those from other fields view it as less open and less ethically reliable. This indicates that disciplinary background can influence perceptions of ChatGPT’s ethical performance and transparency.

\begin{table}[H]
\centering
\begin{threeparttable}
  \caption{Differences in trust-dimension scores: Computer Science vs.\ other fields}
  \label{tab:dim_field}
  \begin{tabular}{lcc}
    \toprule
    Dimension & $U$ & $p$\\
    \midrule
    Expertise        & 1608.5 & .368\\
    Predictability   & 1192.0 & .095\\
    Human-likeness   & 1534.5 & .663\\
    Risk             & 1815.0 & .029$^{*}$\\
    Ease of use      & 1751.0 & .082\\
    Transparency     & 1796.0 & .047$^{*}$\\
    Reputation       & 1503.5 & .810\\
    \bottomrule
  \end{tabular}
  \begin{tablenotes}[para,flushleft]\footnotesize
    * $p<.05$ (Mann–Whitney $U$ tests).
  \end{tablenotes}
\end{threeparttable}
\end{table}

We also examined how trust dimensions vary by LLM familiarity, LLM understanding, and ChatGPT usage frequency (Table~\ref{tab:dim_kruskal}). Significant differences were found primarily in relation to usage frequency. Frequent users reported higher trust across several dimensions, including expertise, predictability, risk, ease of use and transparency. No significant differences were observed for familiarity or understanding. These findings suggest that regular use plays a greater role in shaping trust than general awareness or perceived comprehension. In line with RQ2, this supports the idea that trust is reinforced through hands-on interaction, particularly when users perceive ChatGPT as competent, transparent and user-friendly. While this may seem counterintuitive, since frequent users might also notice limitations, it is possible that repeated positive experiences strengthen trust. Alternatively, those who already trust the tool may be more inclined to use it frequently, pointing to a potential bidirectional relationship between trust and use.

\begin{table}[H]
\centering
\begin{threeparttable}
  \caption{Trust-dimension differences by familiarity, understanding and usage}
  \label{tab:dim_kruskal}
  \begin{tabular}{lccc}
    \toprule
    Dimension & Familiarity $H$ & Understanding $H$ & Usage $H$ ($p$)\\
    \midrule
    Expertise      & 1.84 (.668) & 0.11 (.743) & 8.84 (.003$^{*}$)\\
    Predictability & 1.63 (.687) & 0.02 (.896) & 4.34 (.037$^{*}$)\\
    Human-likeness & 0.73 (.786) & 0.06 (.800) & 0.73 (.393)\\
    Risk           & 2.46 (.620) & 0.07 (.793) & 7.10 (.008$^{*}$)\\
    Ease of use    & 2.35 (.126) & 0.94 (.333) & 22.17 ($<.001^{*}$)\\
    Transparency   & 0.87 (.768) & 0.80 (.372) & 12.12 ($<.001^{*}$)\\
    Reputation     & 0.16 (.689) & 0.02 (.896) & 3.30 (.070)\\
    \bottomrule
  \end{tabular}
  \begin{tablenotes}[para,flushleft]\footnotesize
    * $p<.05$ (Kruskal–Wallis).
  \end{tablenotes}
\end{threeparttable}
\end{table}

\noindent\textbf{Interview Perspectives.}\;The interview data provide deeper insight into how expertise and predictability influence trust. Tom compared ChatGPT’s expertise to that of a university graduate but stressed that consistency is key: “\textit{If it’s more predictable, you’re able to trust it more. If it gives more consistent answers, you can trust it better.}”

Ishaan echoed this, noting that predictable performance builds confidence: “\textit{If you know that for certain things you can expect a certain quality of output, that makes it predictable.}” However, he also described how inconsistency undermined trust: “\textit{Sometimes it can give you correct answers, and sometimes it gives you wrong answers.}”

Although quantitative data showed no significant link between human-likeness and trust, the interviews revealed varied and often negative reactions to this dimension. Tom had mixed views, appreciating the conversational tone in some contexts but critiquing its robotic feel: “\textit{I like it sometimes […] but sometimes it has more information than I need. You can definitely get a sense when something’s written by ChatGPT […] it has just been like very robotic.}”

Fernanda found ChatGPT’s human-like communication unsettling, describing it as “\textit{absolutely terrifying}” compared to simpler assistants like Siri: “\textit{Siri [is] just being clumsy […] ChatGPT is terrifying.}”

Ella similarly viewed human-like features with suspicion: “\textit{Any attempt that it makes to be human-like makes me distrust it […] I feel like I’m being tricked.}”

These responses suggest that while human-likeness may enhance usability for some, it can also provoke discomfort or reduce trust, especially when perceived as inauthentic or manipulative.
\subsection{RQ3: How Does Trust in ChatGPT Vary by Task Type?}

\begin{figure}[H]
  \centering
  \includegraphics[width=.85\linewidth, trim=0 0 0 27, clip]{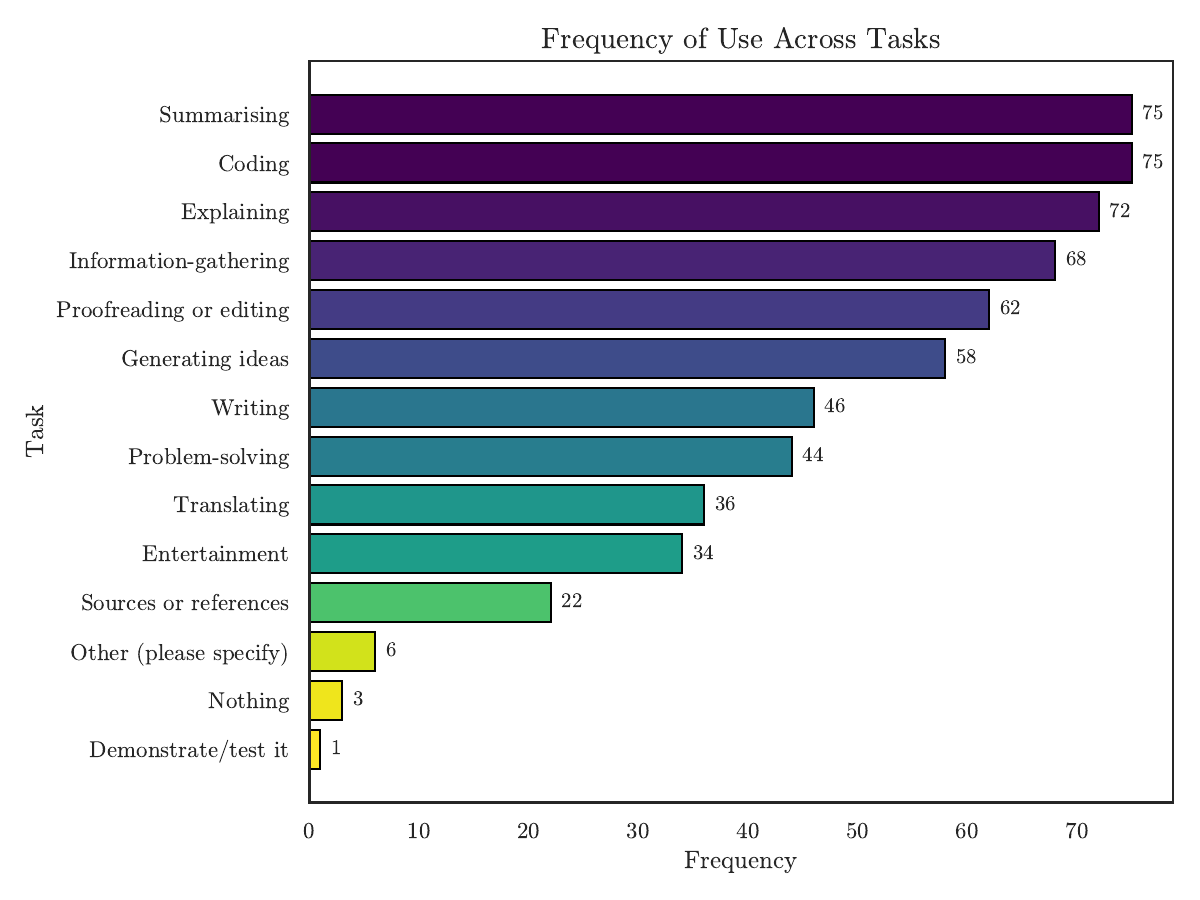}
  \caption{Frequency of ChatGPT use across task types (multiple responses allowed).}
  \label{fig:task_freq}
\end{figure}

Participants indicated whether they had used ChatGPT for specific tasks. As shown in Figure~\ref{fig:task_freq}, summarising and coding were the most frequent uses, reported by about two-thirds of participants. Other common tasks included explaining, information gathering, proofreading and idea generation, each used by over half the sample. Less frequent uses included translating, entertainment and sourcing references.

\begin{figure}[H]
  \centering
  \includegraphics[width=.85\linewidth, trim=0 0 0 27, clip]{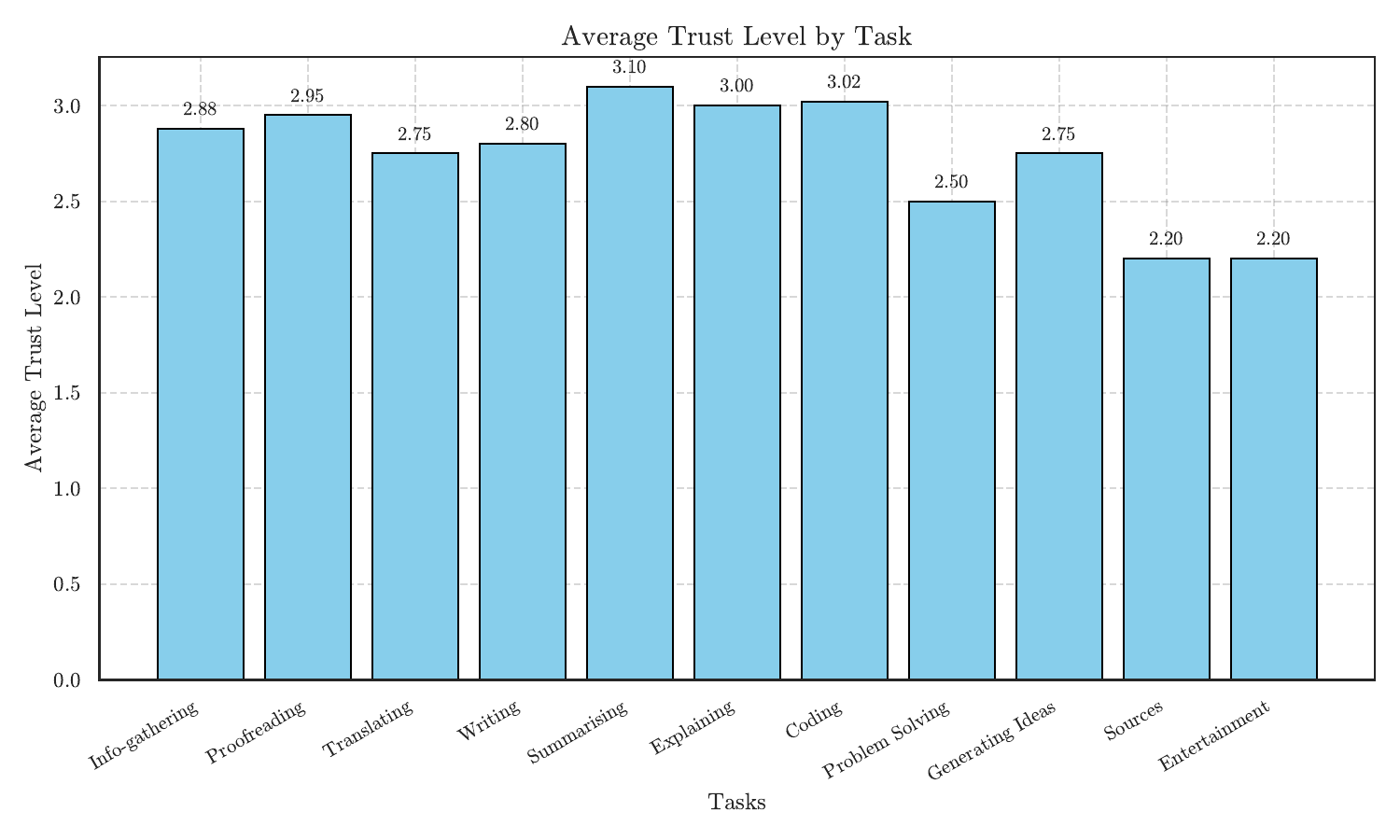}
  \caption{Mean trust in ChatGPT for each task type (1 = not at all, 5 = completely).}
  \label{fig:task_trust}
\end{figure}

Participants then rated their trust in ChatGPT for each task (Figure~\ref{fig:task_trust}). Trust was highest for summarising, coding, and information gathering, while entertainment and idea generation received lower ratings. This pattern suggests that users place more trust in ChatGPT for structured, factual, or outcome-oriented tasks, and less for open-ended or creative ones.

To examine task-specific differences in trust, we conducted post-hoc pairwise comparisons ( Table~\ref{tab:pairwise_tasks}). Trust in ChatGPT was significantly higher for coding tasks than for problem-solving, explanation, and sourcing references. Summarising also received significantly higher trust than sourcing and entertainment tasks. In contrast, sourcing references and entertainment were among the least trusted, with significantly lower ratings than most other tasks. These findings address RQ3 by showing that trust in ChatGPT varies by task, depending on how reliable and useful users perceive it to be in different contexts.

\begin{table}[H]
\centering
\scriptsize
\caption{Post-hoc Wilcoxon pairwise comparisons of task-trust ratings}
\label{tab:pairwise_tasks}
\begin{tabular}{lrrrrrrrrrrr}
\toprule
 & \rotatebox{90}{Coding} & \rotatebox{90}{Entertainment} & \rotatebox{90}{Explaining} & \rotatebox{90}{Generating Ideas} & \rotatebox{90}{Info-Gathering} & \rotatebox{90}{Problem-Solving} & \rotatebox{90}{Editing} & \rotatebox{90}{Sources} & \rotatebox{90}{Summarising} & \rotatebox{90}{Translating} & \rotatebox{90}{Writing} \\
\midrule
Coding & 1.000 & 0.000* & 0.857 & 0.013* & 0.113 & 0.000* & 0.314 & 0.000* & 0.565 & 0.005* & 0.036* \\
Entertainment &  & 1.000 & 0.000* & 0.001* & 0.000* & 0.131 & 0.000* & 0.643 & 0.000* & 0.003* & 0.000* \\
Explaining &  &  & 1.000 & 0.008* & 0.078 & 0.000* & 0.235 & 0.000* & 0.692 & 0.003* & 0.023* \\
Generating Ideas &  &  &  & 1.000 & 0.368 & 0.079 & 0.140 & 0.000* & 0.002* & 0.749 & 0.701 \\
Info-Gathering &  &  &  &  & 1.000 & 0.008* & 0.564 & 0.000* & 0.031* & 0.223 & 0.607 \\
Problem-Solving &  &  &  &  &  & 1.000 & 0.001* & 0.049* & 0.000* & 0.151 & 0.032* \\
Editing &  &  &  &  &  &  & 1.000 & 0.000* & 0.114 & 0.073 & 0.275 \\
Sources &  &  &  &  &  &  &  & 1.000 & 0.000* & 0.001* & 0.000* \\
Summarising &  &  &  &  &  &  &  &  & 1.000 & 0.001* & 0.008* \\
Translating &  &  &  &  &  &  &  &  &  & 1.000 & 0.481 \\
Writing &  &  &  &  &  &  &  &  &  &  & 1.000 \\
\bottomrule
\end{tabular}
\vspace{2mm}
\begin{flushleft}
\footnotesize
\textit{Note.} Bonferroni-adjusted $p$-values from Wilcoxon pairwise tests. * $p < .05$ (two-tailed).
\end{flushleft}
\end{table}
\FloatBarrier

To examine how overall trust in ChatGPT relates to trust across specific tasks, we conducted a correlation analysis between participants’ overall trust and their task-specific trust ratings (Table~\ref{tab:task_corr}). Significant positive correlations were found for problem-solving, information gathering and sourcing references. This suggests that users who trust ChatGPT on more cognitively demanding or fact-based tasks also tend to report higher overall trust. Notably, trust in ChatGPT’s ability to generate sources was the strongest correlate, despite known issues with citation accuracy. One explanation may be that users conflate confident output with factual accuracy, particularly if they are unaware of these limitations. These findings contribute to RQ3 by suggesting that trust is shaped not only by task type, but also by how objective or factual the task appears to users, even when this perception does not reflect ChatGPT’s actual capabilities.

\begin{table}[H]
\centering
\begin{threeparttable}
  \caption{Overall trust vs.\ task-specific trust (Spearman correlations)}
  \label{tab:task_corr}
  \begin{tabular}{lcc}
    \toprule
    Task & Mean (SD) & $\rho$ ($p$)\\
    \midrule
    Info-gathering    & 2.86 (1.00) & .229 (.014)$^{*}$\\
    Editing           & 2.94 (1.02) & .063 (.504)\\
    Translating       & 2.73 (1.02) & .018 (.852)\\
    Writing           & 2.79 (1.05) & .163 (.081)\\
    Summarising       & 3.05 (1.20) & –.042 (.659)\\
    Explaining        & 3.02 (1.06) & .143 (.126)\\
    Coding            & 3.01 (1.06) & .092 (.329)\\
    Problem-solving   & 2.52 (0.92) & .236 (.011)$^{*}$\\
    Idea generation   & 2.73 (1.10) & –.002 (.982)\\
    Sources           & 2.19 (0.98) & .312 (.001)$^{*}$\\
    Entertainment     & 2.22 (1.08) & .132 (.160)\\
    \bottomrule
  \end{tabular}
  \begin{tablenotes}[para,flushleft]\footnotesize
    * $p<.05$.
  \end{tablenotes}
\end{threeparttable}
\end{table}

We compared responses from Computer Science students and participants from other fields to assess whether task-based trust differs by academic background.(Table~\ref{tab:task_field}). Computer Science students reported significantly higher trust in ChatGPT for proofreading/editing and writing tasks. No significant differences were found for other tasks. These results suggest that discipline influences task-specific trust, with Computer Science students expressing more confidence in ChatGPT for language-related tasks. This may reflect greater familiarity with the tool’s strengths or a perception that these tasks are lower risk. These findings contribute to RQ3 by showing how academic background shapes trust in specific applications of ChatGPT.

\begin{table}[H]
\centering
\begin{threeparttable}
  \caption{Task-trust differences: Computer Science vs.\ other fields}
  \label{tab:task_field}
  \begin{tabular}{lcc}
    \toprule
    Task & $U$ & $p$\\
    \midrule
    Info-gathering    & 1423.5 & .810\\
    Editing           & 1877.5 & .010$^{*}$\\
    Translating       & 1580.0 & .478\\
    Writing           & 1905.0 & .007$^{*}$\\
    Summarising       & 1561.0 & .527\\
    Explaining        & 1628.0 & .303\\
    Coding            & 1616.0 & .341\\
    Problem-solving   & 1436.0 & .871\\
    Idea generation   & 1424.0 & .814\\
    Sources           & 1496.5 & .840\\
    Entertainment     & 1458.5 & .980\\
    \bottomrule
  \end{tabular}
  \begin{tablenotes}[para,flushleft]\footnotesize
    * $p<.05$ (Mann–Whitney $U$).
  \end{tablenotes}
\end{threeparttable}
\end{table}

\subsubsection*{Interview Perspectives}

The interview data further illustrate how task-specific trust is shaped by perceived stakes and task type. Ishaan, a Computer Science postgraduate, reported trusting ChatGPT for low-risk tasks such as grammar correction and proofreading, explaining that he used it to \textit{“check grammatical errors or if there is a flaw in the logic […] small details like this.”} He also highlighted ChatGPT’s strength in summarisation, stating, \textit{“I trust it […] for summarisation or creating good, simple concepts and definitions. I think it’s very good at doing those kinds of tasks.”}

Similarly, Ella, a humanities undergraduate, trusted ChatGPT for summarising and generating starting points for ideas, noting, \textit{“I would trust it to basically like summarise a lot of the information that’s out there […] generate a list of ideas […] give me some starting points.”}

Tom, a Computer Science undergraduate, expressed confidence in ChatGPT’s ability to debug code, explaining, \textit{“I wouldn’t use it for anything that would be relatively important to my life […] but debugging, I trust it for because either it gets it right or I still have the bug.”} These insights support the broader finding that users tend to trust ChatGPT more for lower-stakes, structured tasks, such as summarisation or debugging, where the impact is limited and outputs can be easily verified.

When it came to distrust, participants expressed consistent caution around complex or high-stakes tasks. Ella, a humanities undergraduate, noted her reluctance to rely on ChatGPT for fully formed outputs, stating, \textit{“I would never trust it to like give me a finished product.”} Tom, a Computer Science undergraduate, described hesitancy around generating longer code, explaining, \textit{“I wouldn’t trust it for generating large amounts of code […] anything over 30–40 lines of code, it just completely forgets what it’s supposed to be doing.”}

Similarly, Fernanda expressed scepticism about using ChatGPT in educational contexts, particularly for exercises that require nuanced input: \textit{“I wouldn’t trust it for exercises for the classes too […] there are a lot of variables that go into it.”}

Ishaan also voiced concerns about ChatGPT’s reliability in information retrieval, especially for specialised queries: \textit{“For information retrieval […] I wouldn’t trust ChatGPT at all.”} He highlighted its limitations in sourcing and referencing academic material accurately.
\subsection{RQ4: How Do Perceptions of AI’s Societal Impact Relate to Trust?}

\begin{figure}[H]
  \centering
  \includegraphics[width=.8\linewidth,
                   trim=0 30 0 44, 
                   clip]{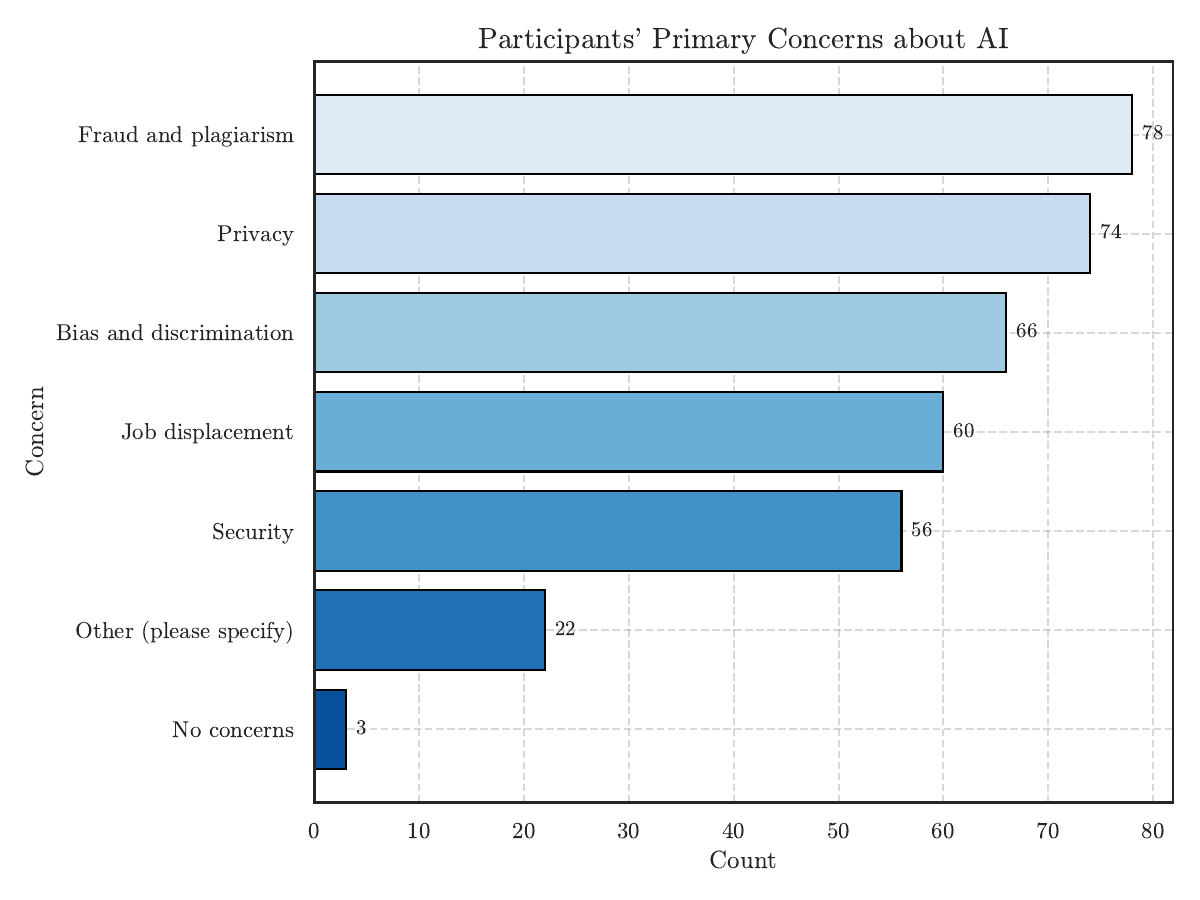}
  \caption{Most frequently cited concerns about AI.}
  \label{fig:ai_concerns}
\end{figure}

Most participants viewed AI’s societal impact as both positive and negative, recognising its potential benefits alongside associated risks. Specifically, 53\% described it as having both positive and negative effects, 38.3\% perceived it as primarily positive, and only 8.7\% saw it as negative. Participants were asked to indicate their concerns about AI by selecting from a list of potential societal risks (Figure~\ref{fig:ai_concerns}). The most common concerns focused on fraud and plagiarism, privacy, and bias or discrimination. Fraud and plagiarism, cited most frequently, reflect anxieties about generative AI misuse, especially in academic and professional settings, highlighting broader challenges around integrity and accountability. Concerns about job displacement and security were also prominent, underscoring worries about automation’s long-term impact on employment and safety. Only two participants reported no concerns, possibly reflecting a higher level of trust in the technology. 

When comparing trust in ChatGPT across perception groups, those who saw AI as predominantly positive reported the highest trust (Table~\ref{tab:ai_impact_trust}). Pairwise Dunn tests 

When comparing trust in ChatGPT across different perceptions of AI’s societal impact, the greatest variation was observed among participants who viewed AI as both positive and negative ($M = 3.02$, $SD = 0.96$). Those who saw AI as predominantly positive reported the highest trust ($M = 3.55$, $SD = 0.73$), while participants with a negative view reported trust levels similar to those with a mixed perspective ($M = 3.10$, $SD = 0.99$). Pairwise comparisons (Table~\ref{tab:ai_impact_trust}) showed a significant difference only between the “Both positive \& negative” and “Positive” groups. These findings suggest that a positive view of AI’s societal impact could relate to higher trust in ChatGPT.

\begin{table}[H]
\centering
\begin{threeparttable}
  \caption{Pairwise comparisons of trust by perception of AI’s impact}
  \label{tab:ai_impact_trust}
  \begin{tabular}{lcr}
    \toprule
    Group comparison & Mean diff. & $p$\\
    \midrule
    Both vs.\ Negative  & 0.08 & .958\\
    Both vs.\ Positive  & 0.53 & .008$^{*}$\\
    Negative vs.\ Positive & 0.45 & .322\\
    \bottomrule
  \end{tabular}
  \begin{tablenotes}[para,flushleft]\footnotesize
    Dunn post-hoc after Kruskal–Wallis $H = 8.39$, $p = .015$; * $p<.05$.
  \end{tablenotes}
\end{threeparttable}
\end{table}

\subsubsection*{Interview Perspectives}
The qualitative data further emphasise participants’ dual perspectives on AI’s societal impact, aligning closely with concerns identified in the survey. Ella, a humanities undergraduate, worried that AI could undermine creativity and intellectual property in the arts, warning it might be used to \textit{``rip off artists.''} Ishaan, a CS postgraduate, raised concerns about job displacement, particularly for low-skilled workers: \textit{``There is a class of people who would be affected […] low-skilled jobs that are repetitive.''} Fernanda, a humanities postgraduate, expressed concern that AI might discourage critical thinking in education: \textit{``It worries me how much students are going to not exercise critical thinking.''}

Participants also highlighted AI’s potential benefits, particularly in improving productivity and accessibility. Ishaan noted that tools like ChatGPT and Grammarly could help users improve their writing and access information more effectively, especially for non-native English speakers: \textit{``It provides you good information […] to gather more relevant information.''} Ella also saw potential for productivity gains, saying AI could \textit{``make us a lot more productive […] focus on the more complicated or even creative stuff.''}

Bias in AI emerged as a major concern across interviews. All participants acknowledged that generative models inherit human biases from their training data. Tom, a CS undergraduate, pointed out that tools like DALL-E tend to reinforce stereotypes: \textit{``If you ask DALL-E to create an image […] chances are it would give an image of a white man.''} He suggested users could mitigate bias through prompt engineering. Ishaan echoed this, observing that ChatGPT’s Western-centric training limits its cultural sensitivity: \textit{``For marginalised communities and people who don’t speak English, it’s not necessarily most perceptive.''} Ella and Fernanda also voiced concern, noting that since AI is trained on biased human content, the tool itself will carry that bias.

Transparency was another recurring theme. One participant criticised the lack of accessible information about ChatGPT’s training data and methods: \textit{``There’s not much information available […] so I’d say there’s fairly limited transparency.''} Another interviewee stressed that understanding how AI arrives at its responses could improve user trust: \textit{``If you could provide reasoning or references […] that would have been very helpful.''} An undergraduate humanities student similarly said her trust would increase if she understood the tool’s inner workings, even if the output wasn’t perfect: \textit{``If I understood the process […] I wouldn’t decrease my trust as much.''} She called for legal regulation requiring transparency and clarity on data sourcing and funding. A postgraduate humanities student went further, expressing existential concerns about AI’s development: \textit{``If we cannot trust humans to do the right thing, how can you guarantee that you can trust an AI created by humans?''} Her comments reflect broader fears about control, accountability and AI’s growing societal role.

\FloatBarrier

\section{Discussion}
\section*{Discussion}

This mixed-methods study clarifies the factors that shape user trust in LLMs such as ChatGPT. By analysing behavioural engagement, task complexity and ethical concerns, we show how users calibrate reliance on generative AI. Participants reported verifying outputs against personal knowledge and external sources, echoing Rowley and Johnson’s trust-triangulation model \cite{rowley2013} and underscoring the need for transparent, verifiable responses. The discussion that follows examines how user background, task-specific expectations, perceived trust dimensions and ethical issues jointly influence trust in ChatGPT.

\subsection*{Factors Influencing User Trust in ChatGPT}

Behavioural engagement emerged as a stronger determinant of trust in ChatGPT than demographic characteristics. Despite earlier work linking disciplinary background to differences in digital literacy and epistemological orientation \cite{shen2023reliability}, the present analysis detected no significant demographic effects. By contrast, usage frequency exhibited a positive association with trust, suggesting that repeated successful interactions reinforce confidence over time. Interview evidence corroborated this interpretation, as participants described an iterative “trial-and-error” process that accords with Hoff and Bashir’s notion of learned trust \cite{hoff2015}. It remains conceivable, however, that trust itself motivates continued use, implying a reciprocal relationship that warrants further investigation.

Familiarity alone did not predict trust. Although prior studies indicate that familiarity combined with technical literacy supports calibrated reliance \cite{bach2024, foehr2020}, the present findings align with research showing that superficial exposure is insufficient to shape trust judgements \cite{elkins2013, klumpp2019}. Future work should examine whether specific constellations, such as high self-reported familiarity coupled with frequent use, provide a more robust predictor of trust.

A particularly salient result concerned technical understanding: respondents with limited knowledge of LLM mechanisms reported higher trust than their technically proficient counterparts. This pattern echoes research on automation bias, wherein users over-rely on ostensibly competent systems when unaware of their limitations \cite{dzindolet2003, madhavan2007, goddard2012}. Interviewees with technical expertise expressed greater scepticism and cited specific deficiencies, whereas novices initially accepted ChatGPT’s fluent output before becoming more critical through experience. These observations suggest that technical insight fosters caution, whereas its absence predisposes users to over-reliance, underscoring the need for user education and transparent system design, particularly in high-stakes contexts.

\subsection*{Trust Dimensions}

Analysing the seven trust dimensions, expertise, predictability, transparency, human-likeness, ease of use, perceived risk (ethical compliance) and reputation, revealed several significant interrelations. Ease of use was inversely related to perceived risk, implying that intuitive interfaces mitigate worries about errors or unintended consequences \cite{dietvorst2020}. Transparency correlated positively with both perceived expertise and ease of use, supporting evidence that clear explanations bolster perceptions of competence and usability \cite{zerilli2022patterns}. In addition, perceived expertise showed a moderate, positive association with predictability, echoing findings that competence signals reinforce expectations of consistent performance \cite{muir1987, madhavan2007}. Taken together, these results confirm that trust in ChatGPT is multidimensional, arising from interacting perceptions of competence, usability, transparency and reliability that develop through iterative user experience.

Group comparisons showed that Computer Science students rated ChatGPT as more transparent and ethically aligned than peers from other disciplines, corroborating evidence that technically trained users are better able to evaluate AI transparency and ethics \cite{zerilli2022patterns, nist2023rmm}. Non-technical participants, by contrast, viewed the system as less transparent and trustworthy, suggesting that disciplinary background shapes ethical as well as competence judgements.

Across the sample, overall trust correlated most strongly with perceived expertise, followed by predictability, ease of use and transparency. Trust rose as the system was judged more competent, dependable, usable and clear. Interviewees likened ChatGPT’s expertise to that of a university graduate yet stressed the need for consistent performance, echoing findings that competence is a primary driver of trust \cite{shen2023reliability, shen2023radiology}, and aligning with NIST’s emphasis on transparency and usability for calibrated reliance \cite{nist2023rmm}.

Reputation also predicted trust, though less strongly, indicating that direct experience outweighs external opinion, consistent with Choudhury and Shamszare’s observation that hands-on interaction is pivotal \cite{choudhury2023}. Regression analysis identified expertise and perceived risk as the strongest predictors, underscoring the importance of functional performance and ethical alignment, and reinforcing evidence that reliability and competence matter more than anthropomorphic traits \cite{nordheim2019, foehr2020}.

Human-likeness itself neither correlated with nor predicted trust. Some participants appreciated a conversational tone, whereas others found it mechanical, echoing Hoff and Bashir’s caution that anthropomorphic cues must be paired with transparency to prevent over-reliance \cite{hoff2015}. Overall, trust in ChatGPT rests more on perceptions of competence, clarity and ethical design than on human resemblance.

\subsection*{Task-Specific Trust}

Trust in ChatGPT proves highly task-contingent. Survey data show that students rely on the model chiefly for summarisation and coding, tasks whose transparent, easily verified outputs inspire confidence, whereas translation and, in particular, reference sourcing remain uncommon, reflecting doubts about bibliographic accuracy \cite{zerilli2022patterns}. Consistent with this pattern, coding and summarising attract the highest trust ratings, while entertainment queries and citation tasks attract the lowest. Yet overall trust rises sharply with confidence in problem-solving, information retrieval and, paradoxically, reference generation, indicating that persuasive fluency can be mistaken for factual reliability \cite{zerilli2022patterns, almarazlopez2023}. The strong association between global trust and perceived citation competence, despite well-documented error rates, illustrates automation bias whereby users over-rely on outputs they cannot readily verify \cite{dzindolet2003, madhavan2007}. Disciplinary effects are modest: computer-science students report higher trust only for proofreading and writing, suggesting that technical expertise sharpens rather than inflates reliance. Overall, these findings underscore that trust is calibrated by task verifiability, perceived objectivity and user expertise, pointing to the need for explicit accuracy cues and targeted user education in academic settings.

\subsection*{Socio-Ethical Considerations}

Participants’ reflections on AI’s broader social consequences were similarly ambivalent. Interviews and survey comments cited fraud, plagiarism, privacy breaches and algorithmic bias, especially in academic and creative work, as threats to integrity and accountability, alongside long-standing anxieties about job displacement and security. These reservations coexist with appreciation for AI’s potential to boost productivity and widen access, illustrating the “optimism-tempered-by-caution” pattern documented in prior research \cite{bach2024, binns2017}. Notably, participants who judged AI’s societal impact as largely positive displayed significantly higher trust in ChatGPT than those who framed it as both positive and negative, suggesting that macro-level attitudes colour micro-level reliance. This linkage underscores the importance of transparent data practices, explicit bias-mitigation strategies and ethically aligned design if generative models are to cultivate responsible trust \cite{shen2023reliability, shen2023radiology, grant2025}.

\section*{Conclusion}

This study examined the factors shaping user trust in AI systems like ChatGPT, focusing on behavioural engagement, trust dimensions, task-specific use and ethical considerations. Trust was found to be multidimensional, influenced by perceptions of competence, ease of use, transparency and ethical alignment. Participants reported higher trust when ChatGPT was seen as predictable and effective, particularly for low-stakes tasks like coding and summarisation. In contrast, trust was lower for high-stakes tasks such as sourcing references.

Concerns about AI’s societal impact were common, including issues like fraud, plagiarism, privacy, bias and job displacement. Participants with a more positive view of AI’s societal role reported higher trust in ChatGPT, suggesting that broader ethical perceptions influence trust beyond personal experience.

Qualitative insights further revealed that users with technical knowledge were more critical of ChatGPT, while less experienced users tended to trust its fluent outputs initially. Over time, both groups adjusted their trust based on experience, highlighting the need for greater system transparency and user education to support informed, calibrated trust.

\section*{Implications for Policy, Teaching and System Design}

These findings have important implications for policy, education and system design. Policymakers and developers should prioritise transparency, particularly around data use, biases and system limitations, to help users better understand and evaluate AI systems like ChatGPT. Incorporating ethical frameworks such as NIST~\cite{nist2023rmm} and AI HLEG~\cite{ec2019} can support trust while ensuring fairness and accountability.

In education, promoting AI literacy is essential. Teaching users, especially those from non-technical backgrounds, how these systems work and what their limitations are can support more critical engagement and calibrated trust. Integrating AI literacy into university curricula across disciplines would help prepare students to use these tools effectively and ethically.

For system design, the results highlight the need for transparent, ethically aligned AI tools that support verification and explainability, especially for high-stakes tasks. Designers should enable users to understand and assess AI outputs and build features that help them detect and manage bias, supporting more responsible and informed AI use.

In conclusion, this study underscores the need to balance user trust in AI with awareness of its societal and ethical implications. Promoting transparency, education and ethical design can help stakeholders foster responsible trust in tools like ChatGPT, supporting their beneficial use while mitigating risks.

\section*{Funding}

This work was funded by UK Research and Innovation (UKRI) under the UK government’s Horizon Europe funding guarantee [grant number 10039436].

\bibliographystyle{unsrt}
\bibliography{references}

\end{document}